\documentclass[aps, prd, twocolumn, preprintnumbers, letterpaper, nofootinbib, superscriptaddress, balancelastpage, 10pt]{revtex4-1}

\usepackage[utf8]{inputenc}
\usepackage{graphicx}
\usepackage{amsmath, amsfonts, amssymb}
\usepackage{bm, bbm}
\usepackage{enumitem}
\usepackage[caption=false]{subfig}
\usepackage{hyperref}
\usepackage{xcolor}

\newcommand{\beq}{\begin{equation}\begin{aligned}{}}
\newcommand{\eeq}{\end{aligned}\end{equation}}
\newcommand{\beqa}[1]{\begin{equation}\begin{aligned}{#1}}
\newcommand{\eeqa}{\end{aligned}\end{equation}}

\newcommand{\fr}[2]{\dfrac{#1}{#2}}

\newcommand{\del}{\partial}
\newcommand{\dd}{\mathrm{d}}

\newcommand{\Mpl}{M_\text{Pl}}

\usepackage{tikz}

\definecolor{lime}{HTML}{A6CE39}
\DeclareRobustCommand{\orcidicon}{\hspace{-1mm}
	\begin{tikzpicture}
	\draw[lime, fill=lime] (0,0)
	circle [radius=0.16]
	node[white] {{\fontfamily{qag}\selectfont \tiny \,ID}};
	\draw[white, fill=white] (-0.0525,0.095)
	circle [radius=0.007];
	\end{tikzpicture}
	\hspace{-3mm}
}

\foreach \x in {A, ..., Z}{\expandafter\xdef\csname orcid\x\endcsname{\noexpand\href{https://orcid.org/\csname orcidauthor\x\endcsname}
			{\noexpand\orcidicon}}
}

\begin{document}

\title{Heavy QCD Axion in $b\to s$ transition: Enhanced Limits and Projections}

\preprint{KEK-TH-2295}

\author{Sabyasachi Chakraborty}
\email{schakraborty5@fsu.edu}
\affiliation{Department of Physics, Florida State University, Tallahassee, FL 32306, USA}

\author{Manfred Kraus}
\email{mkraus@hep.fsu.edu}
\affiliation{Department of Physics, Florida State University, Tallahassee, FL 32306, USA}

\author{Vazha Loladze}
\email{vloladze@fsu.edu}
\affiliation{Department of Physics, Florida State University, Tallahassee, FL 32306, USA}

\author{Takemichi Okui}
\email{tokui@fsu.edu}
\affiliation{Department of Physics, Florida State University, Tallahassee, FL 32306, USA}
\affiliation{Theory Center, High Energy Accelerator Research Organization (KEK), Tsukuba 305-0801, Japan}

\author{Kohsaku Tobioka}
\email{ktobioka@fsu.edu}
\affiliation{Department of Physics, Florida State University, Tallahassee, FL 32306, USA}
\affiliation{Theory Center, High Energy Accelerator Research Organization (KEK), Tsukuba 305-0801, Japan}

\begin{abstract}
We study a ``heavy'' QCD axion whose coupling to the standard model is dominated by $a G \widetilde{G}$ but with $m_a \gg m_\pi f_\pi / f_a$. This is well motivated as it can solve the strong CP problem while evading the axion quality problem. It also poses interesting challenges for its experimental search due to its suppressed couplings to photons and leptons. Such axion with mass around a GeV is kinematically inaccessible or poorly constrained by most experimental probes except B-factories. We study $B \to K a$ transitions as a powerful probe of the heavy QCD axion by performing necessary 2-loop calculations for the first time, together with some improvement on the existing analysis strategy. We find some of the existing limits are enhanced by at least an order of magnitude. We also demonstrate that the bounds are robust against unknown UV physics. For forthcoming data sets of the Belle~II experiment, we provide a projection that $f_a$ of a few TeV is within its future reach, which is relevant to the quality problem.
\end{abstract}

\maketitle


\section{Introduction}
Null signals of new physics at the TeV scale so far suggest us to adopt broader perspectives on the priorities of theoretical questions and future experimental programs. In particular, the possibility of new physics at scales much lighter than the TeV scale has been gaining growing attention. The axion offers a strong motivation for such light new physics, being a compelling solution~\cite{Weinberg:1977ma, Wilczek:1977pj} to the long-standing strong CP problem~\cite{tHooft:1976rip} by utilizing the Peccei-Quinn (PQ) symmetry~\cite{Peccei:1977hh, Peccei:1977ur}, as well as being a candidate for cold dark matter~\cite{Preskill:1982cy, Dine:1982ah, Abbott:1982af}.

The original axion model~\cite{Weinberg:1977ma,Wilczek:1977pj}, in which QCD is the sole origin of the axion mass, predicts the relation $m_a f_a \simeq m_\pi f_\pi$ among the axion mass $m_a$, its decay constant $f_a$, and the analogous quantities for the pion. If we imagine that the PQ symmetry breaking scale, $f_a$, is related to the origin of the electroweak symmetry breaking, it would be natural to place $f_a$ at the TeV scale, as proposed in the original axion models by Weinberg~\cite{Weinberg:1977ma} and Wilczek~\cite{Wilczek:1977pj}, which then puts $m_a$ at the keV scale by the relation above. This possibility, however, is excluded by astrophysical observations~\cite{Anastassopoulos:2017ftl, Raffelt:2006cw, Raffelt:1996wa, Friedland:2012hj} and beam dump experiments~\cite{Bjorken:1988as, Blumlein:1990ay, Bergsma:1985qz}. Much higher $f_a$ around $10^9$--$10^{13}$~GeV, and hence much lighter $m_a$, can be motivated by an axion as cold dark matter~\cite{Kim:2008hd}. This part of the parameter space has also been searched with null results~\cite{Arik:2008mq,Asztalos:2011bm,Arik:2011rx}.

The relation, $m_a f_a \simeq m_\pi f_\pi$, can easily be violated, however, if there are additional contributions to the axion mass~\cite{Dimopoulos:1979pp,Holdom:1982ex,Dine:1986bg,Flynn:1987rs,Choi:1988sy,Rubakov:1997vp, Choi:1998ep, Berezhiani:2000gh,Choi:2003wr,Hook:2014cda,Fukuda:2015ana,Dimopoulos:2016lvn,Agrawal:2017ksf,Agrawal:2017eqm, Gaillard:2018xgk, Hook:2019qoh,Gherghetta:2020keg, Choi:1998ep, Choi:1988sy,Gupta:2020vxb}.
This permits us to reconsider the case where $f_a$ is at or moderately above the TeV scale, but now with $m_a$ much heavier than $\sim m_\pi f_\pi / f_a \sim$~keV\@. (It would be difficult, if not impossible, to imagine a scenario where $m_a$ is \emph{lighter} than $\sim m_\pi f_\pi / f_a$.) It is particularly important to explore the masses of $10~{\rm MeV}\lesssim m_a$ with TeV-scale $f_a$~\cite{Mariotti:2017vtv}. Such low values of $f_a$ can also be motivated theoretically as a solution to the axion quality problem. If the violation of global symmetries by quantum gravity appears as unsuppressed $O(1)$ coefficients times powers of $f_a / \Mpl$, the axion solution to the strong CP problem would be ruined~\cite{Kamionkowski:1992mf,Holman:1992us,Barr:1992qq,Ghigna:1992iv} \emph{unless} $f_a$ is below $\sim 10$~TeV\@~\cite{Agrawal:2017ksf}.

In this work, therefore, we focus on what we call the \emph{heavy QCD axion} scenario, where $m_a$ is much heavier than $\sim m_\pi f_\pi / f_a$ and the axion couples to the SM dominantly via only the $a G_{\mu\nu} \widetilde{G}^{\mu\nu}$ interaction, where $a$ is the axion field and $G_{\mu\nu}$ the gluon field strength.
Our effective Lagrangian thus has the form
\begin{align}
\label{treelevellagrangian}
\mathcal{L}
=\mathcal{L}_\text{SM}
+\frac{\alpha_s}{8\pi} \frac{a}{f_a} G^{a}_{\mu\nu} \widetilde{G}^{a \mu\nu}
+\frac{1}{2} (\del_\mu a)^2 - \frac{m_a^2}{2} a^2
\,,
\end{align}
where $\widetilde{G}^{a \mu\nu} \equiv \frac12 \epsilon^{\mu\nu\rho\sigma} G^a_{\rho\sigma}$.
The additional terms required for renormalization that are phenomenologically relevant will be discussed shortly.
There are many models that UV-complete this EFT or could do so with minor modifications~\cite{Fukuda:2015ana, Agrawal:2017eqm, Agrawal:2017ksf, Gaillard:2018xgk, Gherghetta:2020keg,Gupta:2020vxb}.

The status of experimental probes into the heavy QCD axion is the following. For $m_a\lesssim 400$~MeV, $f_a$ at the TeV scale is excluded by the hadronic production and diphoton decay of the heavy axion, the proton beam dump experiment \cite{Bergsma:1985qz, Aloni:2018vki}, the kaon experiments \cite{Georgi:1986df, Bardeen:1986yb, Alves:2017avw, Gori:2020xvq, Artamonov:2005ru, Ceccucci:2014oza, Abouzaid:2008xm, Ahn:2018mvc, Bauer:2021wjo},  the precision measurement of pion decay \cite{Aguilar-Arevalo:2019owf,Pocanic:2003pf, Altmannshofer:2019yji}, the fixed target experiment \cite{AlGhoul:2017nbp, Aloni:2018vki, Aloni:2019ruo}, and the collider experiments \cite{Abbiendi:2002je, Knapen:2016moh, Aloni:2018vki}.
 For $m_a\gtrsim 400~{\rm MeV} $, the search is difficult because the hadronic decay mode dominates, and it is overall poorly explored until $m_a$ reaches 50~GeV where the CMS dijet search kicks in \cite{Sirunyan:2017nvi, Mariotti:2017vtv}. However, axion production from hadron decays such as $\phi\to\gamma a$ and $\eta' \to \pi\pi a$ constrain some parameter space \cite{Tanabashi:2018oca} (see also Fig.~\ref{fig:moneyplot}).
A part of the experimental loophole, $m_a \gtrsim 2~\text{GeV}$, can be explored at the LHC if the axion can decay into diphotons \cite{CidVidal:2018blh, CidVidal:2018eel}.

We thus see that the heavy QCD axion with $m_a$ in the few GeV range and $f_a$ at the TeV scale and above has not been explored. In this region, $B$ physics should play a crucial role, having the right energy scale as well as great experimental precision. Moreover, the experimental reach of $B$ physics will be improved further in upcoming years by LHCb (300 fb$^{-1}$) and Belle~II ($5\times 10^{10}$ $B$-meson pairs). A promising channel is $B\to Ka$ with the axion subsequently decaying to hadrons, which is induced at 2-loop %
\footnote{If there is an $aW\tilde{W}$ coupling, $b \to s a$ is induced at one-loop level~\cite{Izaguirre:2016dfi} (see also \cite{Gavela:2019wzg}).},
starting from the tree-level Lagrangian~(\ref{treelevellagrangian}).
The importance of this channel was pointed out in \cite{Aloni:2018vki,Bauer:2020jbp}, but the required 2-loop calculation has not been performed to date. The previous work~\cite{Aloni:2018vki} relies on order of magnitude estimation for axion production.

Our goal, therefore, is to perform this calculation and obtain robust and competitive bounds for the heavy QCD axion. We will also provide a projection for the reach of Belle~II.

\section{Calculation of $b\to s a$}

\begin{figure}[t]
\includegraphics[width=0.45\textwidth]{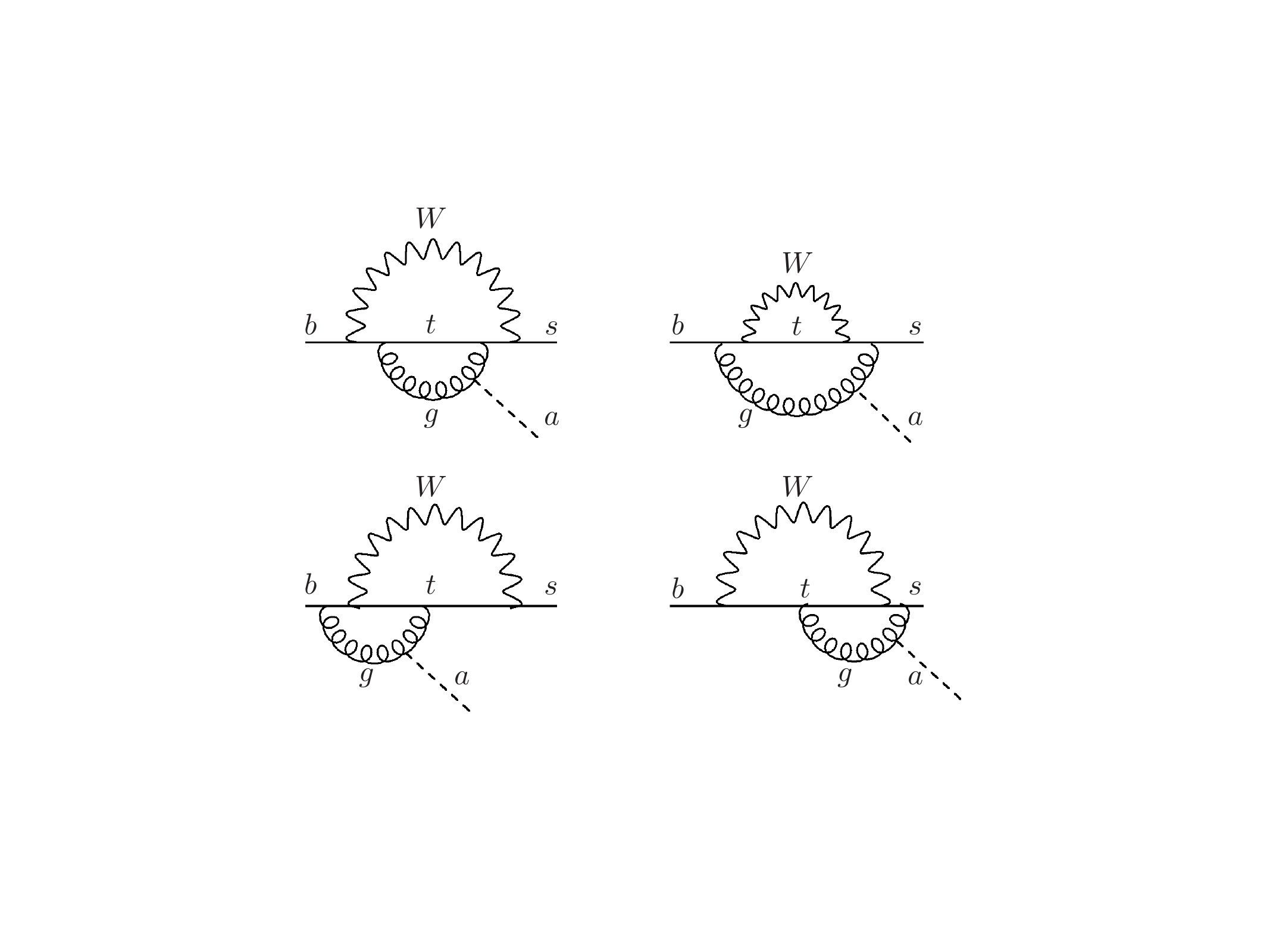}
\caption{Leading 1-particle-irreducible diagrams for $b \to sa$ from the Lagrangian~(\ref{treelevellagrangian}).}
\label{fig:2-loop-diagrams}
\end{figure}

\begin{figure}[t]
\includegraphics[width=0.20\textwidth]{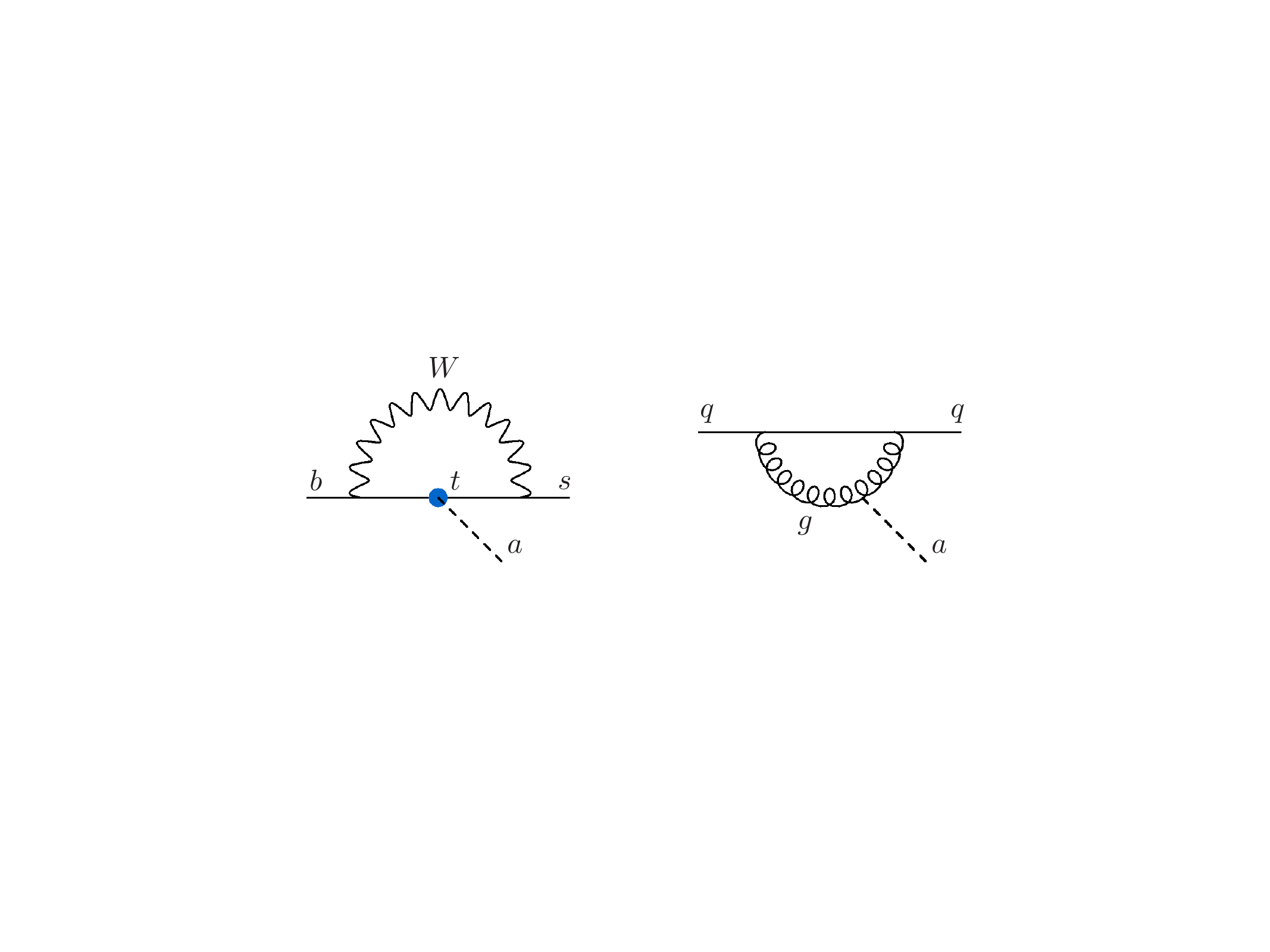}
\caption{The diagram that generates the $C_{qq}$ term of Eq.~(\ref{counterterms}).}
\label{fig:counter-term-diagram}
\end{figure}

Starting from the Lagrangian~(\ref{treelevellagrangian}),
the leading contribution to $b\rightarrow sa$ arises at 2-loop as shown in Fig.~\ref{fig:2-loop-diagrams}.
Cancelling UV divergences in these diagrams requires the following additional interactions to be further included in the Lagrangian:

\begin{align}
\mathcal{L}
= \cdots +
C_{qq} \sum_{q} \frac{\partial_\mu a}{f_a} \,
\bar{q} \gamma^\mu\gamma_5 q
+ C_{bs} \frac{\partial_\mu a}{f_a} \,
\bar{s}_\text{\tiny L} \gamma^\mu\gamma_5 b_\text{\tiny L}
+\text{h.c.},
\label{counterterms}
\end{align}
where the ellipses denote the terms in Eq.~(\ref{treelevellagrangian}) as well as those irrelevant for the $b \to sa$ phenomenology of our interest
(see e.g.~\cite{Bauer:2017ris,Chala:2020wvs} for those other operators generated at 1-loop from Eq.~(\ref{treelevellagrangian})).
The $C_{qq}$ term is generated at 1-loop from the diagram shown in Fig.~\ref{fig:counter-term-diagram}
and necessary to cancel 1-loop sub-divergences in Fig.~\ref{fig:2-loop-diagrams}.
The $C_{bs}$ term is required to remove remaining divergences at 2-loop.
We have written the same coefficient $C_{qq}$ for all quark flavors
because we assume $m_t/\Lambda_{\text{UV}} \ll 1$ and ignore corrections of order $\sim m_t^2/\Lambda_{\text{UV}}^2$,
where $\Lambda_\text{UV}$ is the cutoff of our EFT\@.

It is not necessary at the 2-loop level to modify the coefficient of $a G\widetilde{G}$ in Eq.~(\ref{treelevellagrangian}) from $\alpha_s/8\pi f_a$, provided that the $\alpha_s$ here is treated as the running coupling $\alpha_s(\mu)$. While this claim is verified by an explicit calculation in Appendix, it may be understood as follows. If we treat the axion as an external field, the coefficient of $(a/f_a) G\widetilde{G}$ is completely fixed by matching the PQ-QCD-QCD anomaly. All corrections from turning $a$ back on as a dynamical field involve the $a G \widetilde{G}$ coupling itself at least twice and hence negligibly small.


Although $C_{qq}$ and $C_{bs}$ are free parameters in the EFT,
their sizes must be consistent with the defining feature of our framework that the $a G \widetilde{G}$ interaction is the dominant coupling of the axion to the SM\@.
As we would set $C_{qq}$ and $C_{bs}$ to zero for our scenario if there were no UV divergences requiring their presence as counter-terms,
we regard them as having sizes roughly similar to the coefficients of the corresponding divergences (i.e., those of the $1 / \epsilon$ poles in dimensional regularization).
We thus take $C_{qq} \sim C_F (\alpha_s / 4\pi) (g_s^2 / 16\pi^2) = C_F (\alpha_s / 4\pi)^2$ (see Fig.~\ref{fig:counter-term-diagram}) with $C_F = 4/3$.
For $C_{bs}$, we further include two electroweak gauge couplings and GIM suppression (see Fig.~\ref{fig:2-loop-diagrams}),
so $C_{bs} \sim C_F (\alpha_s / 4\pi)^2 (\alpha_{w}/4\pi) \sum_k V_{kb} V^\ast_{ks} \xi_k$,
where $V$ is the CKM matrix and $\xi_k \equiv m_k^2 / M_W^2$ with $k=u,c,t$.
Therefore, at the cutoff $\Lambda_\text{UV}$ of our EFT, where it is matched on to the UV theory, we parametrize $C_{qq}$ and $C_{bs}$ as
\beq
C_{qq}(\Lambda_{\text{UV}})
&\equiv A C_F
\biggl( \frac{\alpha_s}{4\pi} \biggr)^{\!\! 2}
\,,\\
C_{bs}(\Lambda_{\text{UV}})
&\equiv B C_F
\biggl( \frac{\alpha_s}{4\pi} \biggr)^{\!\! 2}
\frac{\alpha_w}{4\pi}
\sum_kV_{ik}V^\ast_{kj}\xi_k
\,,
\label{cthreeinitial}
\eeq
where $A$ and $B$ are $O(1)$ parameters that depends on the unknown UV completion of the Lagrangian in Eq.~(\ref{treelevellagrangian}). All the SM parameters are evaluated at $\Lambda_{\text UV}$.
We will show, however, that our bounds on $f_a$ are fairly insensitive to the exact values of $A$ and $B$.
Then, keeping in mind these rough sizes of $C_{qq}$ and $C_{bs}$,
we find the leading RG running of $C_{qq}$ and $C_{bs}$
(see Appendix for the details of the calculation):
\begin{align}
\mu\frac{\dd C_{qq}}{\dd\mu}
&= -6 C_F
\biggl( \frac{\alpha_s}{4\pi} \biggr)^{\!\! 2},
\label{ctworun}\\
\mu\frac{\dd C_{bs}}{\dd\mu}
&=
\biggl(
3 C_F \biggl( \frac{\alpha_s}{4\pi} \biggr)^{\!\! 2}
+ C_{qq}
\biggr)
\frac{\alpha_w}{4\pi} \sum_k\xi_kV_{kb}V^\ast_{ks}
\,.\label{cthreerun}
\end{align}

\begin{figure}[h]
\includegraphics[width=0.45\textwidth]{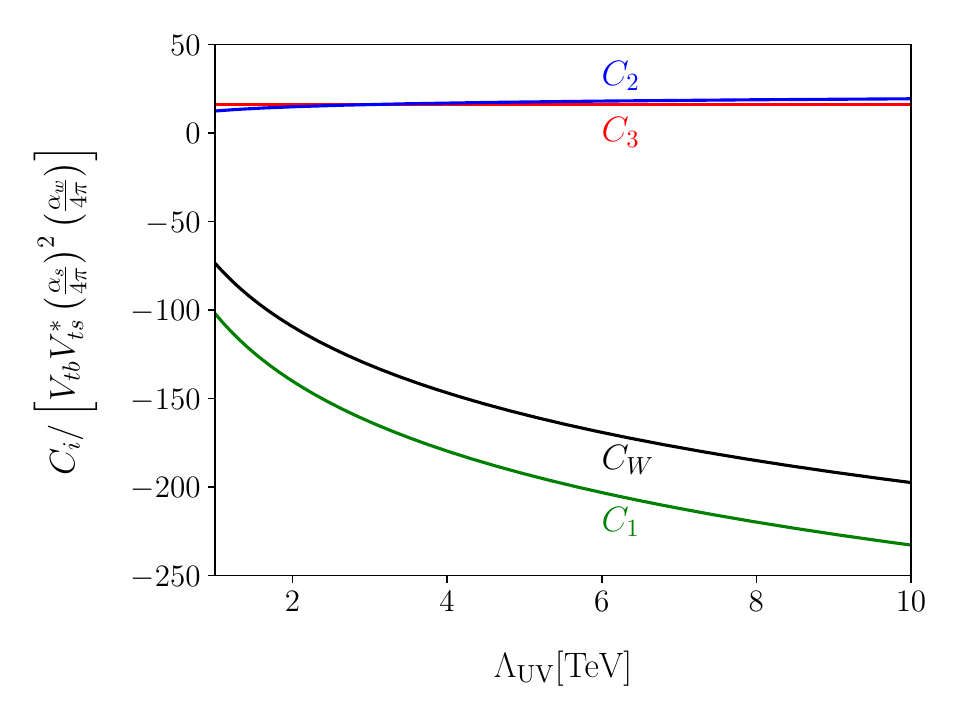}
\caption{$C_1$, $C_2$ and $C_3$ refers to the first, second and third contribution in $C_W$ respectively, for different UV scales (see Eq.~(\ref{wilsonfinal})).}
\label{fig:relative_sizes}
\end{figure}
\begin{figure}[h!]
\centering
{
	\includegraphics[width=0.5\textwidth]{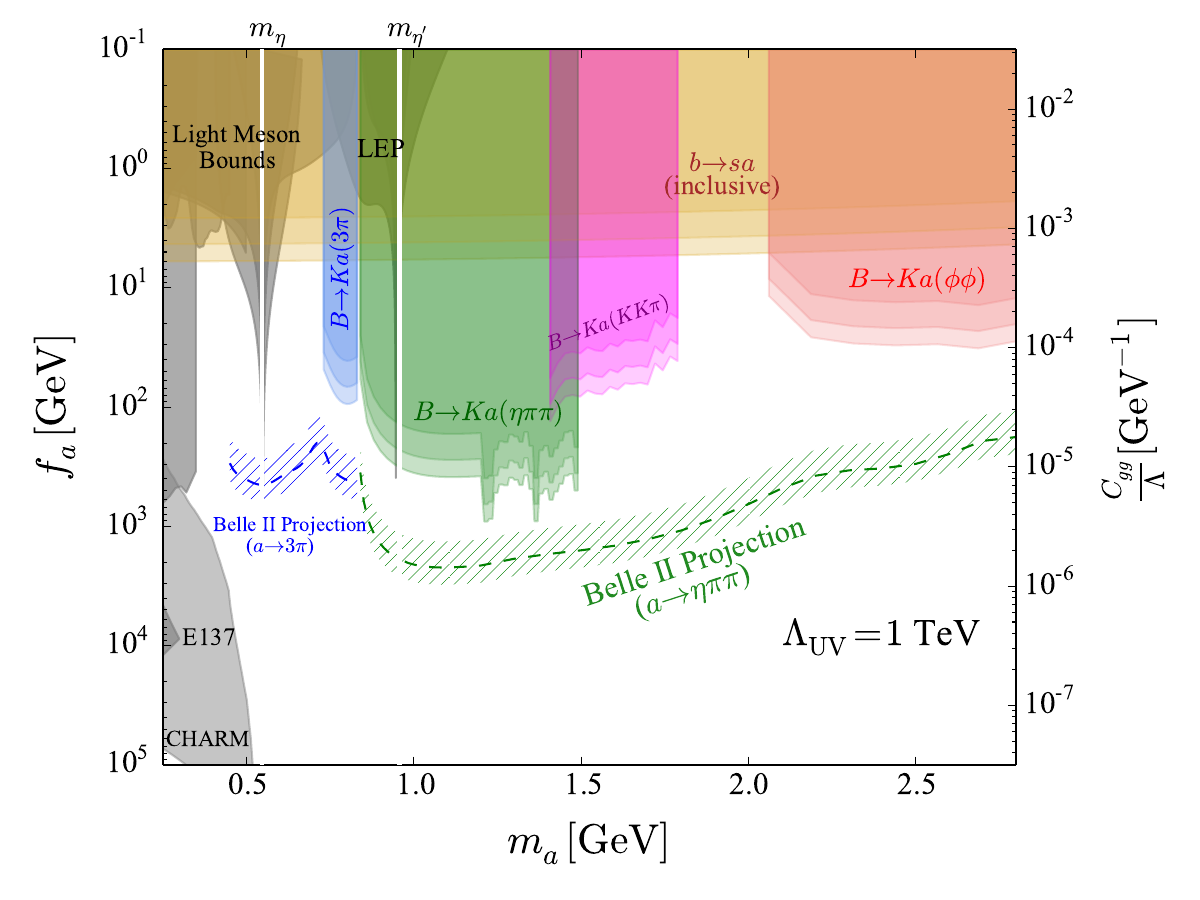} }
{
	\includegraphics[width=0.5\textwidth]{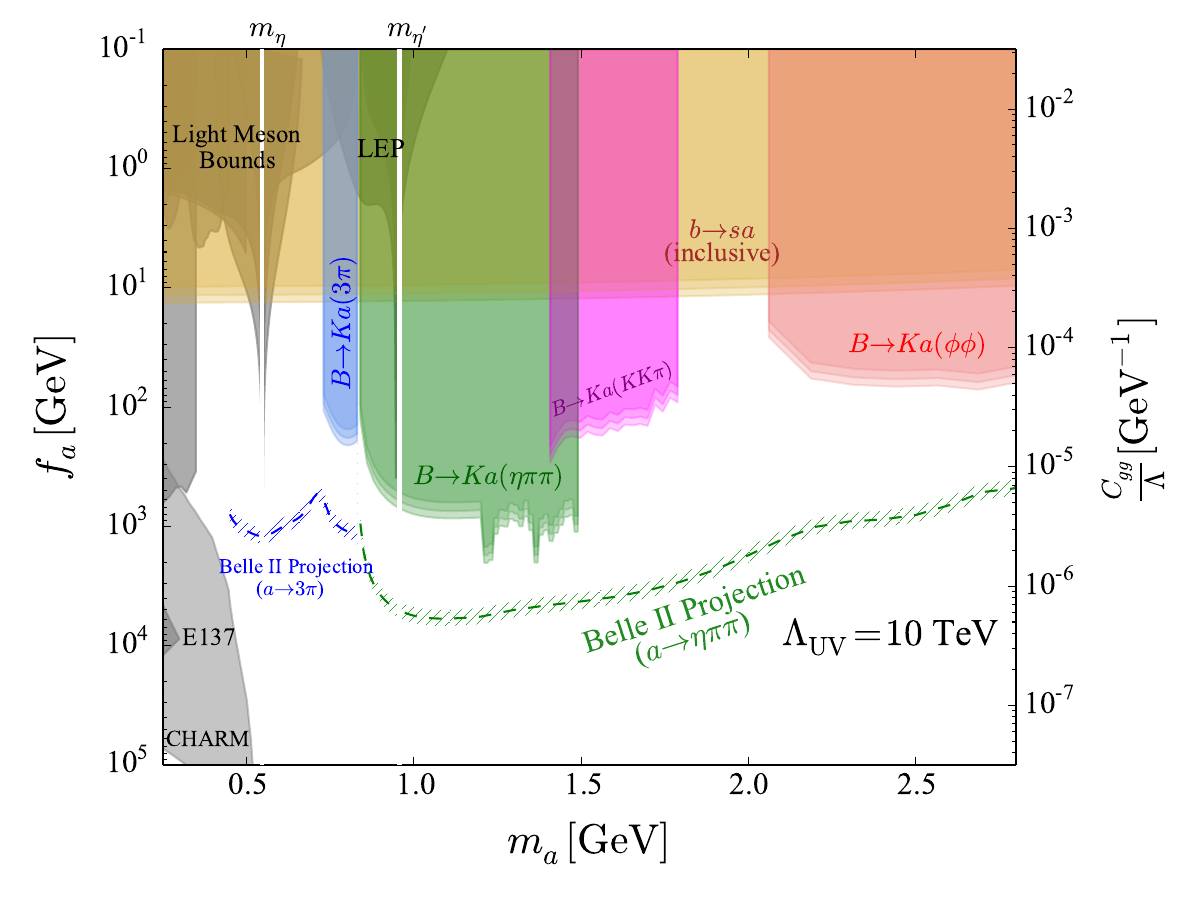} }
\caption{We portray the constraints from different $B$-decay measurements in the $m_a$-$f_a$ plane. Three curves are drawn for each constraint corresponding to different initial conditions (see Eq.\eqref{cthreeinitial}), i.e., the strongest  $\left(A=+3, B=-3\right)$, weakest $\left(A=-3, B=+3\right)$ and central constraints $\left(A=B=0\right)$. We choose the UV scale $\Lambda_{\text{UV}}$ to be 1 and 10 TeV for the top and the bottom plot, respectively. See main texts about systematic uncertainties from the form factor of Eq.\eqref{eq:formfactor} and the data-driven calculation of the axion branching fractions.
The grey shaded regions comprise bounds from \cite{Bergsma:1985qz, Bjorken:1988as, Aloni:2018vki, Aloni:2019ruo, Gori:2020xvq, Abbiendi:2002je, Knapen:2016moh, Bauer:2021wjo}.  For $B\to K a$, we use \cite{Zyla:2020zbs} for inclusive analsysis and \cite{Chobanova:2013ddr, Lees:2011zh, Aubert:2008bk} for exclusive channels $a\to 3\pi, \eta\pi\pi, KK\pi, \phi\phi$. For the projection at Belle~II (dashed lines), $5\times 10^{10}$ $\bar{B}B$ pair is assumed, and the band shows the dependence on the different initial conditions. The right vertical axis is labelled using the notation of Ref.~\cite{Aloni:2018vki} for comparison.}
\label{fig:moneyplot}
\end{figure}

After running down to $\mu \sim M_W$ using Eqs.~(\ref{ctworun}, \ref{cthreerun}),
we switch to another EFT in which the top quark and $W$ boson are integrated out.
In the limit of $m_{b,s} / M_W \to 0$,
this new EFT contains only one operator relevant for the $b \to sa$ phenomenology:
\beq
\mathcal L_{bsa}
=
C_W
\frac{\partial_\mu a}{f_a}
\bar{s}_\text{\tiny L} \gamma^\mu\gamma_5 b_\text{\tiny L}
+\text{h.c.}
\,,\label{matchingoperator}
\eeq
where $C_W$ is determined by $C_{qq}(\mu_w)$ and $C_{bs}(\mu_w)$ with $\mu_w \sim M_W$ and the contributions from integrating out $t$ and $W$.
We find
\beq
C_W
= C_{bs}(\mu_w)
+\frac{\alpha_w}{4\pi} \, C_{qq}(\mu_w) \, g(\mu_w)
+\frac12
\frac{\alpha_w}{4\pi}
\biggl( \frac{\alpha_s}{4\pi} \biggr)^{\!\! 2}
f(\mu_w)
\,,\label{wilsonfinal}
\eeq
where $g$ and $f$ are 1- and 2-loop matching functions given respectively in Eqs.~(\ref{gfunction}, \ref{ffunction}) in Appendix.
In the limit of $m_{b,s} / M_W\rightarrow 0$, $C_W$ does not run between $M_W$ to $m_b$. This is because in this particular limit, there is no mixing between $aG\tilde{G}$ and flavor changing axial-vector coupling. In Fig.~\ref{fig:relative_sizes} we show the 1st, 2nd, and 3rd terms of the right-hand-side of Eq.~(\ref{wilsonfinal}) as well as the net $C_W$, all as a function of $\Lambda_\text{UV}$, assuming the initial condition $A=B=0$ in Eq.~(\ref{cthreeinitial}). We observe that $C_{bs}$, i.e., the $b$-$s$-$a$ operator dominates the overall $C_W$ and interferes destructively with $C_{gg}$, i.e., $a$-$g$-$g$ operator. The dominance of $C_{bs}$ can be explained by the operator mixing under the RGE evolution; $C_{bs}$ acquires leading logarithmic contributions $\sim \ln(\Lambda_\text{UV}^2/M_W^2)$ and $\sim \ln^2(\Lambda_\text{UV}^2/M_W^2)$ due to the mixing with $a$-$g$-$g$ and $a$-$q$-$q$ operators. Since $\ln({\rm TeV}^2/M_W^2)\approx 5$ is a relatively large number, $C_{bs}$ dictates over others.

The final step is to evaluate the meson level decay  $B\rightarrow aK^{(*)}$ \cite{Batell:2009jf,Izaguirre:2016dfi}. We find
\beq
\Gamma_{B\rightarrow Ka}
= \bigl| C_W \bigr|^2 \frac{m_B^3}{64\pi f_a^2}
\!\left( 1 - \frac{m^2_K}{m^2_B} \right)^{\!\! 2} \!
\lambda_{Ka} \,
\bigl[ f_0(m^2_a) \bigr]^2
\,,
\eeq
where $\lambda_{Ka}$ is given by
\beq
\lambda_{Ka}
=\!\left[
\!\left( 1 - \frac{(m_K+m_a)^2}{m^2_B} \right)\!
\!\left(1-\frac{(m_K-m_a)^2}{m^2_B} \right)\!
\right]^{\!\frac12}\!,
\eeq
while $f_0(m_a^2)$ is the form factor obtained from the light-cone QCD sum rules \cite{Ball:2004ye,Ball:2004rg}:
\begin{eqnarray}
f_0(m_a^2) = \frac{0.330}{1-m_a^2/37.5\>\text{GeV}^2}\label{eq:formfactor}
\,.\label{eq:formfactorf0}
\end{eqnarray}
It is important to note that these form factors derived from QCD sum rules have $O(10)\%$ uncertainties~\cite{Ball:2004ye,Ball:2004rg}. The approximate branching ratio is given by ${\rm BR}(B^+\to K^+ a)\approx 1.1 (7.6)\times 10^{-5}[f_a/100~{\rm GeV}]^2$ for $\Lambda_{\rm UV}=1 (10)~{\rm TeV}, A=B=0$ and $m_a=1~{\rm GeV}$ (the  dependence of $m_a$ is smaller than 10\% in the parameter space we consider).

\section{Phenomenology}
To derive constraints on the axion decay constant as a function of the mass we use different $B$ decay measurements.
\begin{itemize}[leftmargin=*]
 \item We first derive the constraint on inclusive $b\rightarrow sa$ decay based on  PDG data $\text{BR}(B^+\rightarrow \bar{c} X)=97\pm4 \%$ \cite{Zyla:2020zbs}. Thus, we  require $\text{BR}(b\rightarrow sa)<1-\text{BR}(b\rightarrow c)\lesssim 11\% $. Note that this constraint does not contain any uncertainties coming from hadronization or calculation of axion decay.  Therefore this is most robust bound derived in this paper. For  inclusive branching fraction we  use:
\begin{eqnarray}\label{inclusivebranching}
\text{BR}(b \to sa)
&\simeq&
\frac{\bigl| C_W \bigr|^2}{\Gamma_B f_a^2}
\frac{(m_B^2-m_a^2)^2}{32\pi m_B}
\;,
\end{eqnarray}
where $\Gamma_B$ is the width of  $B$ meson.
The inclusive $b\to s a$ decay rules out the region marked by yellow in Fig.~\ref{fig:moneyplot}. In fact, this constraint is comparable and in some cases more robust than the bounds drawn for light meson phenomenology~\cite{Aloni:2018vki, Gori:2020xvq}, e.g., $K_L\to\pi^0 a\left(\gamma\gamma\right)$,  $\eta^\prime\to\pi\pi a\left(3\pi\right)$, $\phi\to\gamma a\left(\pi\pi\gamma,\eta\pi\pi\right)$ and $\gamma p\to p a\left(\gamma\gamma\right)$,  displayed in grey in Fig.~\ref{fig:moneyplot}.

 \item Next we use exclusive final states $a\to3\pi$, $\phi\phi$, $KK\pi$, and $\eta\pi\pi$ to perform axion search.  We perform a peak search except in $a\to3\pi$ final state.
 To calculate corresponding branching fractions for axion decay we use the data-driven approach given in Ref.~\cite{Aloni:2018vki} and use branching fractions given in Fig.~3 of their paper.
The uncertainties in this approach for axion hadronic (partial) widths are not estimated in~\cite{Aloni:2018vki} so they are not included in the following bounds. However, these can be extracted by the same drive-driven method of Ref.~\cite{Aloni:2018vki}.
\end{itemize}
\begin{enumerate}[leftmargin=*]
        \item The constraints on the $a\rightarrow 3\pi$ channel, shown by the blue region in Fig.~\ref{fig:moneyplot} is  derived based on Belle analysis \cite{Chobanova:2013ddr}. This analysis is applicable to $0.73~{\rm GeV} \leq m_a\leq 0.83$ GeV.
        We require $\text{BR}\left(B^{0}\to K^{0}a\right)\text{BR}\left(a\to\pi^+\pi^-\pi^0\right)<4.9\times 10^{-6}$, which is from $\text{BR}\left(B^0\to K^0 \omega\right)<5.5\times 10^{-6}$ \cite{Chobanova:2013ddr} and $\text{BR}\left(\omega \to\pi^+\pi^-\pi^0\right)=89\%$.

        \item We use $B\rightarrow K\phi\phi$ data of BaBar \cite{Lees:2011zh}  to  derive a constraint on the $a\rightarrow \phi \phi$ channel, which is shown by the orange region in Fig.~\ref{fig:moneyplot}.
        We assume the axion to be at the center of each bin (see Fig. 5 of Ref.~\cite{Lees:2011zh}) of width 125 MeV. Despite experimental smearing, the gaussian event distribution coming from the axion decay is expected to be completely inside one of these bins. From the perspective of peak search, we also require the signal from the axion to be less than the central value of the measurement augmented with 2$\sigma$ uncertainty.

        \item We analyze  $B\rightarrow K a(\rightarrow K K\pi)$ final state based on Babar measurements \cite{Aubert:2008bk}. The channel is studied at LHCb using 3fb$^{-1}$ data \cite{Aaij:2016xas}, but the sensitivity is currently weaker compared to Babar.
        The bound is shown by the pink region in Fig.~\ref{fig:moneyplot}.  To derive this bound, we follow a similar strategy mentioned previously with one difference. The bin size for $KK\pi$ experimental data is only 22.5 MeV (see Fig. 1(e) of Ref.~\cite{Aubert:2008bk}). Hence, instead of assuming the axion mass to be at the center of each bin, we assume it to be at the boundary of adjacent bins. We then require the number of events from the decay of the axion to be less than the sum of central values of those two bins plus $2\sigma$ uncertainty, after subtracting non-resonant background from the measurement. The merging of two bins correct for any spilling over effect due to experimental smearing.
     Further, experimental efficiency is calculated based on binned data and final measurement of the branching fraction given on Fig.~1 (e) and TABLE I of \cite{Aubert:2008bk} respectively.  Finally, the data analysis performed on $K K\pi$ measurement contains mass cut: one of the $K\pi$ pair invariant  mass is required to be $0.85~\text{GeV}\lesssim m_{K\pi} \lesssim 0.95~\text{GeV}$. To apply this cut on axion decay calculations we use $a\rightarrow K K\pi$ matrix element given in Eq.~(S59-S61) of \cite{Aloni:2018vki}. However, the result strongly depends on the experimental input parameters that have large uncertainties. Because of this uncertainties bound from this channel have order one error close to the end of the mass spectrum $m_a\sim1.8~\text{GeV}$.

        \item For $a\rightarrow \eta\pi\pi$ \cite{Aubert:2008bk} in the $1.2~ \text{GeV}<m_a<1.5~\text{GeV}$ window, we do everything similarly to $KK\pi$ except the mass cut. For $m_a<1.2~\text{GeV}$ one can notice that the number of measured events are less than for $m_a>1.2~\text{GeV}$. Therefore, we take the weakest constraint from $m_a>1.2~\text{GeV}$ region and extend this branching ratio bound for low axion mass $m_a<1.2~\text{GeV}$. As depicted by the green shaded region in Fig.~\ref{fig:moneyplot}, this channel provides the strongest constraint on the parameter space.
\end{enumerate}
\begin{itemize}[leftmargin=*]
  \item Finally we derive Belle~II projection  for $a\rightarrow \eta\pi\pi$ and $a\rightarrow 3\pi$ search, shown as the green and blue dashed curves in Fig.~\ref{fig:moneyplot}.
 \end{itemize}
 \begin{enumerate}[leftmargin=*]
 \item To estimate $a\rightarrow \eta\pi\pi$ projection we first extrapolate BaBar's continuous QCD background given on FIG.1  (f) of~\cite{Aubert:2008bk}. Next, we scale it  with luminosity, assuming $5\times 10^{10}$ $\bar{B}B$ pair at Belle~II. Eventually, based on our result we calculate standard deviation and require that signal  from  the axion to be less than 2 times this standard deviation. We estimate the experimental resolution of the axion mass as $\delta m_a\sim\delta m_{\eta\pi\pi} m_a/m_{\eta'}$ where $\delta m_{\eta\pi\pi} \sim 13.4$ $\text{MeV}$ is the experimental resolution of the $\eta'$ mass fitted from the Fig.1  (f) of~\cite{Aubert:2008bk}.

\item To derive $a\rightarrow 3\pi$ projection we do everything similarly to $a\rightarrow \eta\pi\pi$ except we use Fig.~2 (d) of \cite{Chobanova:2013ddr}, which shows background in $0.73 ~\text{GeV}<m_a<0.83~\text{GeV}$ range. We assume background for $m_a<0.73~\text{GeV}$ to be same as at $m_a=0.73~\text{GeV}$. Also, we use fixed $\delta m_a\approx40~\text{MeV}$ experimental resolution of the axion mass, that is estimated using the signal shape of $\omega$ shown on the same figure.
\end{enumerate}
Finally, it is a nontrivial result of our study that the bounds on $f_a$ are not very
sensitive to the exact values of A and B, mainly because the double
logarithmic enhancement we calculated in this paper dominate. This means that the bounds and projections obtained in our paper are independent of the exact nature of unknown UV physics. 
Moreover, since the uncertainties in the form factors~(\ref{eq:formfactorf0}) are $O(10)\%$, so our calculation also shows that the detail of UV physics is relatively a minor effect, particularly for large UV scales.

\section{Conclusion}
\begin{figure}[h]
\includegraphics[width=0.5\textwidth]{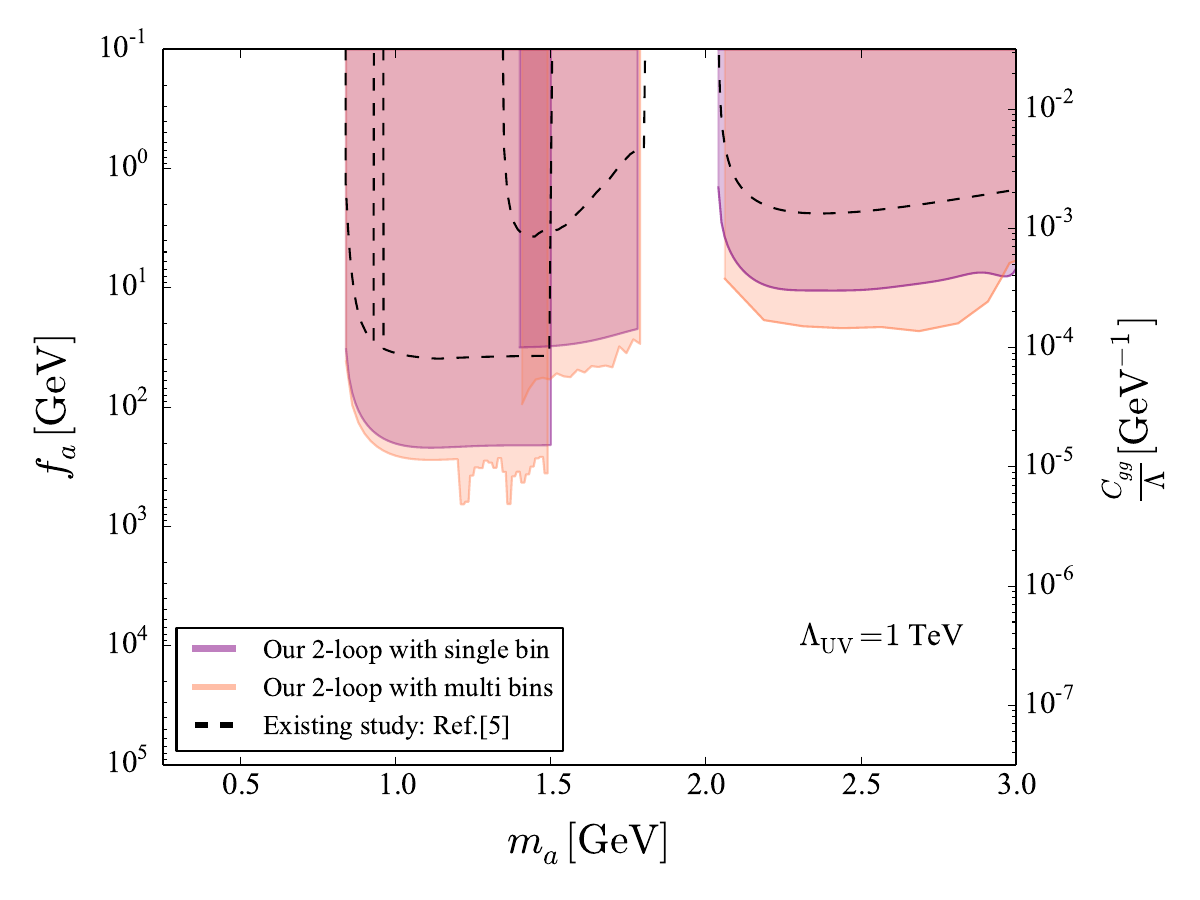}
\caption{In this figure, the dark red colored regions are obtained with our two loop amplitude. This shows a factor of 5 or 6 enhancement from the axion production rate compared to the previous estimation obtained in~\cite{Aloni:2018vki} (shown by the dashed lines). Both of these results use single bin analysis as mentioned in the text. Additional enhancement of factor 2, shown by the light red region, comes from our multiple bin analysis.}
\label{fig:compare}
\end{figure}

In this letter, we  performed the first 2-loop calculation for the axion production from $B\to K a$ process starting from the minimal interaction of the QCD axion, $a G\tilde{G}$ (Eq.\ref{treelevellagrangian}). Assuming the UV scale to be at 1 TeV, the constraints on the $m_a$-$f_a$ parameter space (see Fig.~\ref{fig:moneyplot}) turns out to be $\displaystyle{\sim10}$ times stronger than the previous estimate~\cite{Aloni:2018vki}. Increasing the UV scale only increases this difference. The reason for this enhancement is two fold. Firstly, in~\cite{Aloni:2018vki} the amplitude was a rough estimate up to $O(1)$ factors in the coefficients and without the complete logarithmic enhancement.
In contrast, our work, for the first time, provides the complete leading 2-loop amplitude with RG improvement for the axion-induced FCNC processes, which exhibits an enhancement by a factor of about 5 or 6 in the axion production rate.
Secondly, we perform a detailed bin by bin analysis instead of assuming an overall branching fraction. This makes our bound even more robust by roughly a factor of two at least and sometimes more as shown in fig.~\ref{fig:compare}. 
Combining these two effects, our bounds are enhanced by at least an order of magnitude with respect to the previous rough estimates in Ref.~\cite{Aloni:2018vki}. Therefore, the bounds on the decay constant is order of 100 GeV using Belle and BaBar measurements for $\Lambda_{\text{UV}}=1$ TeV.
For the future, although there are many intensive studies for the heavy QCD axion based on the (near) future data at kaon factories \cite{Gori:2020xvq}, GlueX \cite{Aloni:2019ruo}, LHC with track-trigger \cite{Hook:2019qoh, Gershtein:2020mwi}, DUNE near detector \cite{Kelly:2020dda}, or beam-dump type facilities (summarized in Fig.41 of \cite{Beacham:2019nyx}),
the $B\to K a$ process is particularly important for GeV mass range of the axion. This is because the GeV axion is not produced at light meson precision experiments and also because the lifetime is shorter due to the hadronic decay channelsng the beam-dump experiments less effective.
Belle II will be able to cover the unique parameter space using $B\to K a (\to \eta\pi\pi,3\pi)$ as shown in Fig.~\ref{fig:moneyplot}, and we expect the other channels and future data of LHCb will further improve the sensitivities. Also, $B\to K a (\to \gamma\gamma)$ will be another attractive channel particularly for $m_a<3m_\pi\simeq 450$~MeV, which is not yet studied in B-factories.

\acknowledgments
We thank Mike Williams and Yotam Soreq for correspondence regarding Ref.~\cite{Aloni:2018vki}. We also thank  Lina Alasfar, Fernando Febres Cordero, Andrei Gritsan, TaeHyun Jung and Abner Soffer for discussions. This work is supported by by the US Department of Energy grant DE-SC0010102.

\appendix

\section{Renormalization Scheme}
We start from the EFT Lagrangian:
\beq
\mathcal L =
\mathcal L_\text{SM} + \mathcal{L}_{a} +
\sum_i C_i \mathcal O_i
+\ldots
\,,\label{eq:app:lagrangian1}
\eeq
where $\mathcal{L}_a$ denotes the axion kinetic and potential terms and the ellipses represent effective operators irrelevant for the $b \to sa$ phenomenology of our interest,
while $i \in \{gg,qq,bs\}$ and
\beq
& \mathcal O_{gg} = \frac{1}{8\pi} \frac{a}{f_a}
G^{a}_{\mu\nu}\tilde{G}^{ a \mu\nu}\;, \\
& \mathcal O_{qq} = \sum_{q}\frac{\partial_\mu a}{f_a} \,
\bar{q}\gamma^\mu\gamma_5q\;, \\
& \mathcal O_{bs} = \frac{\partial_\mu a}{f_a} \,
\bar{s}_\text{\tiny L}\gamma^\mu\gamma_5 b_\text{\tiny L}
+\text{h.c.}
\,.\label{eq:app:operators}
\eeq
However, we will soon be redefining $\mathcal{O}_{qq}$ and $\mathcal{O}_{bs}$ below in order to take into account the subtleties of dealing with $\gamma_5$ in dimensional regularization (DR)\@.

To simplify our calculations, we will neglect terms of order $m_{b,s,a}^2 / M_W^2$ or higher.
This in particular means that we evaluate the diagrams in Fig.~\ref{fig:2-loop-diagrams} at vanishing external momenta.
The Feynman gauge has been used throughout our calculations and thus the inclusion of an unphysical Nambu-Goldstone mode accompanying every $W$ boson is implied in the following discussions.
We have implemented tensor reduction in FORM~\cite{Vermaseren:2000nd} and used KIRA~\cite{Maierhoefer:2017hyi} to obtain integration-by-parts relations.

We will regulate UV divergences using DR,
while we cut off IR divergences explicitly by introducing fictitious quark masses.
Note that all diagrams in Fig.~\ref{fig:2-loop-diagrams}
as well as all coefficients in Eq.~(\ref{wilsonfinal})
are $O(\alpha_s^2 \alpha_w)$.
At this order, the dependence on the fictitious masses actually cancels out as the IR theory~(\ref{matchingoperator}) has no IR divergences even in the $m_{b,s} \to 0$ limit at the same order.
We have checked this cancellation explicitly as a validation of our calculations.

The absence of anomalous chiral fermion loops in the diagrams of Fig.~\ref{fig:2-loop-diagrams} allows us to adopt the following simple prescription for $\gamma_5$ and $\epsilon^{\mu\nu\rho\sigma}$.
We first redefine $\mathcal{O}_{qq}$ and $\mathcal{O}_{bs}$ as
\beq
& \mathcal O_{qq}
=\frac{{\rm i}}{6}
\frac{\partial_\mu a}{f_a}
\epsilon^{\mu\nu\rho\sigma}
\sum_{q}
\bar{q} \gamma_\nu\gamma_\rho\gamma_\sigma q
\,,\\
& \mathcal O_{bs}
=\frac{{\rm i}}{6}
\frac{\partial_\mu a}{f_a}
\epsilon^{\mu\nu\rho\sigma}
\bar{s}_\text{\tiny L} \gamma_\nu\gamma_\rho\gamma_\sigma b_\text{\tiny L}
+\text{h.c.}
\,,\label{eq:app:Othreenew}
\eeq
which is equivalent to their original forms in $d=4$ but we use these new forms in $d = 4-2\epsilon$ because what we directly obtain from diagrams in Fig.~\ref{fig:2-loop-diagrams} is actually the product of three $\gamma$ matrices multiplied by the $\epsilon$ tensor from the $a$-$g$-$g$ vertex.
Therefore, all we need is the total antisymmetric property of the $\epsilon$ tensor, which we assume as part of the definition of our scheme, and the property $\{\gamma_5, \gamma^\mu\} = 0$, which is valid as we have no anomalous chiral fermion loops.
We do not use any explicit form of the $\epsilon$ tensor nor any relation between $\gamma_5$ and the $\epsilon$ tensor,
until only after all divergences are cancelled and we are back to $d=4$.

We employ the $\overline{MS}$ scheme (with one exception mentioned below) and redefine the Wilson coefficients as
\begin{align}
C_i \to \sum_j (e^{\gamma_{\tiny E}}\mu^2/4\pi)^{\epsilon/2} C_j \gamma_{j i}
\,,\quad
\mathcal O_i \to \sum_j \mathcal{Z}_{i j}\mathcal O_j
\,,\label{redefinitions}
\end{align}
where $\mathcal{Z}$ consists of the field-strength renormalizations of the SM fields inside $\mathcal{O}_i$.
For our 2-loop computation depicted in Fig.~\ref{fig:2-loop-diagrams}, it only has three nontrivial components:
\beq
\mathcal{Z} = \left( \begin{matrix} Z_G & 0 & 0\\ 0 & Z_{q} & Z_{bs} \\ 0 & 0 & 1 \end{matrix} \right)\;,
\label{eq:app:fieldstrength}
\eeq
where $Z_G$ and $Z_q$ are respectively the gluon and quark field-strength renormalizations due to 1-loop QCD corrections, while $Z_{bs}$ the renormalization of the $b$-$s$ kinetic mixing induced by a $W$ loop.
We use $\overline{MS}$ to determine $Z_G$ and $Z_q$, while we fix $Z_{bs}$ by requiring that the net $b$-$s$ kinetic mixing should vanish at 1-loop at vanishing quark momentum.

All of these are determined completely by the SM and we find
\beq
& Z_G = 1+\frac{\alpha_s}{4\pi}\!\left(\frac{5}{3}N_c-\frac{2}{3}N_f\right)\!\frac{1}{\epsilon}\;, \\
& Z_q = 1-\frac{\alpha_s}{4\pi}\frac{C_F}{\epsilon}\;, \\
& Z_{bs} = -\frac{\alpha_w}{4\pi}\sum_k\frac{\xi_k V_{kb} V^\ast_{ks}}{4}
 \bigg[\frac{1}{\epsilon} -\ln\frac{M_W^2}{\mu^2} + \frac{3(\xi_k+1)}{2(\xi_k-1)}\\
& \qquad\qquad\qquad\qquad\qquad\qquad - \frac{\xi_k(2+\xi_k)}{(\xi_k-1)^2}\ln\xi_k\bigg]\;,
    \label{Zbs}
\eeq
where $N_c = 3$ and $N_f=6$.

\section{Anomalous dimensions and renormalization group evolutions}

To obtain the anomalous dimensions matrix $\gamma$ in Eq.~(\ref{redefinitions}), we calculate 1- and 2-loop diagrams contributing to the $a$-$g$-$g$, $a$-$q$-$q$ and $a$-$b$-$s$ vertex corrections in $\overline{MS}$.
We find
\beq
\gamma =
\begin{pmatrix} 1 - \frac{\alpha_s}{4\pi} \frac{\beta_0}{\epsilon}
& -\frac{\alpha_s}{16\pi^2} \frac{3C_F}{\epsilon}
& \frac{\alpha_s}{16\pi^2} \frac{\alpha_w}{4\pi} \frac{3C_F S}{4} \!\left( \frac{1}{\epsilon} - \frac{1}{\epsilon^2} \right)\! \\
0 & 1 & \frac{\alpha_w}{4\pi} \frac{S}{2\epsilon}
\\ 0 & 0 & 1
\end{pmatrix},
\label{Zanomalous}
\eeq
where $\beta_0 = 11 N_c/3 - 2 N_f/3$ and
$S = \sum_k \xi_k V_{kb} V^\ast_{ks}$.
The RGEs can then be found by demanding $\mu\, \dd (\mu^\epsilon C_i \gamma_{i j}) / \dd\mu =0$, i.e.,
\beq
\mu\frac{\dd C_i}{\dd\mu}=-\epsilon C_i-\sum_{j, k}C_j \, \mu\frac{\dd \gamma_{j k}}{\dd\mu}(\gamma^{-1})_{k i}\;.
\eeq
In the limit $\epsilon\rightarrow 0$, we get
\beq
\label{cthreerun1}
&\mu\frac{\dd C_{gg}}{\dd\mu} =-\beta_0\frac{\alpha_s}{2\pi} C_{gg}\;,\\
&\mu\frac{\dd C_{qq}}{\dd\mu} =-6C_F\frac{\alpha_s}{4\pi}\fr{C_{gg}}{4\pi}\;, \\
&\mu\frac{\dd C_{bs}}{\dd\mu} =\biggl(3\frac{\alpha_s}{4\pi}\frac{C_{gg}}{4\pi}C_F+C_{qq}\biggr)\frac{\alpha_w}{4\pi}\sum_k\xi_kV_{kb}V^\ast_{ks}\;.
&
\eeq
Here, to see the size of each contribution, recall that roughly $C_{gg} \sim \alpha_s$ and $C_{qq} \sim (\alpha_s / 4\pi)^2$.
We further simply the RGE for $C_{bs}$ by neglecting the $u$ and $c$ quark masses.
This then allows us to combine $\alpha_w$ and $\xi_t$ as $\alpha_w \xi_t = y_t^2/2\pi$.
Therefore, we also incorporate the SM running of the top-quark Yukawa coupling:
\beq
\mu\frac{\dd y_t}{\dd\mu} \simeq
\frac{y_t}{16\pi^2}
\!\left( \frac{9}{2} y_t^2 - 8g_3^2 \right).
\eeq
We also take into account the running of $V_{ts}$.
The leading contribution reads~\cite{Balzereit:1998id}
\beq
\mu\frac{\dd V_{ts}}{\dd\mu}\simeq\frac{3}{32\pi^2}y_t^2V_{ts}\;.
\eeq

Let us first verify our claim in the main text that the running of $C_{gg}$ is completely accounted for by the SM running of $\alpha_s$.
This can be trivially seen by solveing Eq.~(\ref{cthreerun1}) with the initial condition $C_{gg}(\Lambda_\text{UV})=\alpha_s(\Lambda_\text{UV})$, which leads to $C_{gg}(\mu)=\alpha_s(\mu)$.
Then, setting $C_{gg}(\mu)=\alpha_s(\mu)$ in the RGEs for $C_{qq}$ and $C_{bs}$ above,
we obtain the results in Eq.~(\ref{cthreeinitial}).

After we run from $\Lambda_\text{UV}$ down to $\mu \sim M_W$, we integrate out the $W$ and $t$ and match onto the EFT described by the operator~(\ref{matchingoperator})
with the coefficient~(\ref{wilsonfinal}),
where we find
\begin{widetext}
\beq
\label{ffunction}
\begin{split}
f(\mu) = \frac{3}{2} C_F \sum_k V_{kb} V^\ast_{ks}
\biggl[&
-\frac{3\xi_k}{2} \ln^{2\!}\frac{M_W^2}{\mu^2}
+\biggl\{
 \frac{\xi_k (3\xi_k - 2) (3\xi_k + 4)}{2(\xi_k - 1)^2}
-\frac{(\xi_k - 2) (3\xi_k + 1)}{\xi_k - 1} \ln(\xi_k - 1)
\biggr\} \ln\xi_k
\\
&+\frac{3\xi_k^3-14\xi_k^2-8\xi_k+4}{2(\xi_k-1)^2}\ln^{2\!}\xi_k
+\biggl\{
 \frac{9\xi_k (\xi_k+1)}{2(\xi_k-1)}
-\frac{\xi_k (3\xi_k^2-2\xi_k+8)}{(\xi_k-1)^2} \ln\xi_k
\biggr\} \ln\frac{M_W^2}{\mu^2}
\\
&+\frac{\pi^2(4+11\xi_k-7\xi_k^2)+3\xi_k(13\xi_k-3)}{12(\xi_k-1)}
+\frac{(\xi_k-2)(3\xi_k+1)}{\xi_k-1} \, \text{Li}_2\!\left(\frac{1}{\xi_k}\right)\!
\\
&-\frac{(\xi_k+2)(\xi_k^2+2\xi_k-1)}{(\xi_k-1)^2} \, \text{Li}_2\!\left(\frac{\xi_k-1}{\xi_k}\right)
\biggr]\,.
\end{split}
\eeq

\beq
\label{gfunction}
g(\mu)=
\frac14 \sum_k V_{kb} V^\ast_{ks} \xi_k
\biggl[ \frac{\xi_k+5}{1-\xi_k}
+\frac{2(\xi_k^2-2\xi_k+4)}{(\xi_k-1)^2} \ln\xi_k
+2\ln\frac{M_W^2}{\mu^2}\biggr]
\,.
\eeq
\end{widetext}
We verified that the difference between Eq.~(\ref{gfunction}) and the 1-loop matching function found in Eq.~(71) of Ref.~\cite{Bauer:2020jbp} is due to different scheme choices of handling $\gamma_5$ and the Levi-Cevita tensor.

\section{Input parameters}
\begin{table}[ht]
\centering
\begin{tabular}{|c | c|}
\hline
Parameters & Values \\
\hline\hline
$G_F$ \cite{Tanabashi:2018oca}  & $1.166\times 10^{-5}~\text{GeV}^{-2}$ \\ 
$\alpha_s\left(M_Z\right)$ \cite{Tanabashi:2018oca} & $0.1181\pm 0.002$\\
$V_{tb}$ \cite{Tanabashi:2018oca} & 0.9991 \\
$V_{ts}$ \cite{Tanabashi:2018oca} & 0.0413 \\
$M_W$ \cite{Tanabashi:2018oca} & 80.379 GeV \\
$M_Z$ \cite{Tanabashi:2018oca} & 91.187 GeV \\
$\overline{m_t} \left(m_t\right)$ \cite{Marquard:2015qpa} & $163.6$ GeV \\
$\overline{m_b} \left(m_b\right)$ \cite{Tanabashi:2018oca} & $4.18$ GeV \\
$\overline{m_s} \left(m_s\right)$ \cite{Tanabashi:2018oca} & $92.9$ MeV \\
$m_B$ \cite{Tanabashi:2018oca} & 5.279 GeV \\
$m_K^\pm$ \cite{Tanabashi:2018oca} & 493.6 MeV \\
$m_K^0$ \cite{Tanabashi:2018oca} & 497.6 MeV \\
\hline
\end{tabular}
\caption{Input parameters. }
\label{table:1}
\end{table}
In Table~\ref{table:1} we list the input parameters used in our analysis.

\twocolumngrid
\bibliographystyle{utphys}
\bibliography{references}
\end{document}